\documentclass{emulateapj}
\shorttitle{Spin-orbit alignment}
\shortauthors{Dawson}
\def\prob{{\rm~prob}}
\def\teff{{~T_{\rm eff}}}
\def\kep{\emph{Kepler\ }}
\def\tcut{{~T_{\rm cut}}}
\def\tdec{{\tau_{\rm~decay}}}
\def\tbrake{{\tau_{\rm~brake}}}
\def\alphab{{\alpha_{\rm~brake}}}
\def\talign{{\tau_{\rm~align}}}
\def\tage{{\tau_{\star,\rm~age}}}
\def\ieff{I_{\star,{\rm~eff}}}
\def\oms{\Omega_\star}
\def\mret{M_{p,\rm retro}}
\newcommand{\icarus}{Icarus\ }
\begin{document}
\slugcomment{Submitted to ApJL on April 10, 2014. Accepted on June 6, 2014.}
\title{On the tidal origin of hot Jupiter stellar obliquity trends}
\author{Rebekah I. Dawson\altaffilmark{1}}
\affil{Department of Astronomy, University of California, Berkeley, Hearst Field Annex, Berkeley CA 94720-3411}
\altaffiltext{1}{{\tt  rdawson@berkeley.edu; Miller Fellow}}

\begin{abstract}
It is debated whether the two hot Jupiter populations --- those on orbits misaligned from their host star's spin axis and those well-aligned --- result from two migration channels or from two tidal realignment regimes. Here I demonstrate that equilibrium tides raised by a planet on its star can account for three observed spin-orbit alignment trends: the aligned orbits of hot Jupiters orbiting cool stars, the planetary mass cut-off for retrograde planets, and the stratification by planet mass of cool host stars' rotation frequencies. The first trend can be caused by strong versus weak magnetic braking (the Kraft break), rather than realignment of the star's convective envelope versus the entire star. The second trend can result from a small effective stellar moment of inertia participating in the tidal realignment in hot stars, enabling massive retrograde planets to partially realign to become prograde. The third trend is attributable to higher mass planets more effectively counteracting braking to spin up their stars. Both hot and cool star require a small effective stellar moment of inertia participating in the tidal realignment, e.g., an outer layer weakly coupled to the interior. I demonstrate via Monte Carlo that this model can match the observed trends and distributions of sky-projected misalignments and stellar rotation frequencies. I discuss implications for inferring hot Jupiter migration mechanisms from obliquities, emphasizing that even hot stars do not constitute a pristine sample.
\end{abstract}


\section{Introduction}
The distribution of host star obliquities $\psi$ --- the angle between a planet's orbital angular momentum and its host star's spin angular momentum --- constrains hot Jupiters' origin. Hot Jupiters are thought to form at several AU \citep{Raf}, reaching orbital periods $P$ of several days via: a) high eccentricity migration, in which a Jupiter's initially highly eccentric orbit shrinks and circularizes because of tidal dissipation in the planet, or b) disk migration (e.g., \citealt{Gold}). Mechanisms for the former produce a broad $\psi$ distribution (e.g., \citealt{FT,Naoz11,Chatterjee}), whereas the latter preserves $\psi=0$ (e.g., \citealt{Bitsch}), unless the disk or star becomes misaligned \citep{Tremaine,Batygin,Rogers12, Lai14}. Rossiter-McLaughlin measurements of $\lambda$ \citep{Ross,McL} ($\psi$ sky-projected) could distinguish the mechanism of hot Jupiter migration \citep{MJ,Naoz12} but $\psi$ is sculpted also by tidal dissipation in the star, through which the planet transfers angular momentum from its orbit to the star's spin. 

From early obliquity measurements, \citet{FW} (FW09 hereafter) inferred two hot Jupiter populations: well-aligned and isotropic. Subsequent discoveries linked them to host stars' properties: hosts of misaligned Jupiters have effective temperature $\teff>6250$K (\citealt{W10},W10 hereafter) and $M>1.2M_\sun$ \citep{KS}. Among hosts with $M>1.2M_\sun$, those older than 2-2.5 Gyr are aligned, the age at which such stars develop a significant convective envelope \citep{2013T}. W10 proposed that tidal dissipation is more efficient in cool stars with thick convective envelopes, allowing realignment. \citet{A12} (A12 hereafter) confirmed the temperature break with a larger sample, constructed a tidal dissipation parameter, and demonstrated that misalignment is correlated with that parameter. To allow the realignment to occur on timescale shorter than the planet's orbital decay, W10 suggested that the convective envelopes of cool stars may be sufficiently decoupled from the radiative interior to be realigned separately. Even without stronger dissipation, this decoupling would result in a much shorter timescale for the realignment of cool stars than hot stars.

In the W10 framework, only cool stars experience realignment of their convective envelopes. However, hot Jupiters may also influence the outer layers of hot stars. Hot stars have convective envelopes, which are thinner and therefore arguably easier to realign. For example, a hot Jupiter has synchronized hot star ($\teff=6387$K) $\tau$ Bootis, accomplishable if the star has a thin convective envelope weakly coupled to the interior \citep{Catala}. A second obliquity trend, the mass cut-off for retrograde planets (e.g., \citealt{Hebrard}), may further be evidence that hot stars are not immune to their hot Jupiters' tidal influence. Furthermore, attempts to reproduce the observed trends have resulted in major inconsistencies, such as a missing population of $\psi=180^\circ$ planets or too many oblique cool stars (e.g., \citealt{Lai12,Rogers13, Xue}), although individual systems are modeled successfully \citep{VR,2012H} and $\lambda$ correlates with the theoretical tidal realignment timescale (A12,\citealt{2012H}).
\begin{figure*}[h]
\includegraphics[width=\textwidth]{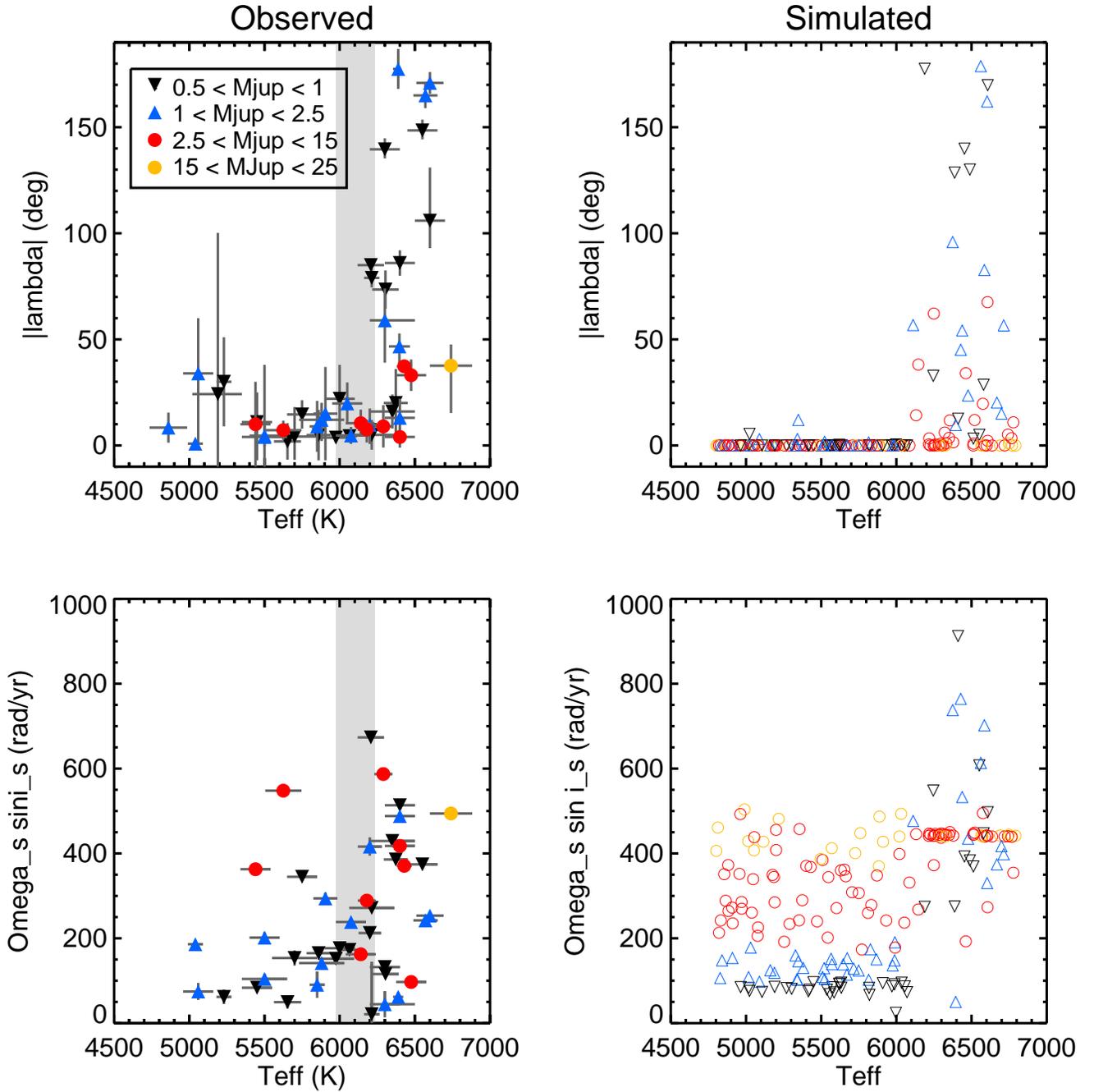}
\caption{Left: Observed sky-projected spin-orbit alignment $(\lambda)$ and rotation frequencies $\oms=v\sin~i/R_\star$ (\citealt{Wright}, A12). Right: Simulated population (\S4). \label{fig:obs}}
\end{figure*}
Here I reconsider the cause of the observed trends. In \S2, I summarize the observations, quantifying the temperature cut-off and linking it to the onset of magnetic braking, which I identify from the \citet{Mc} sample of \kep rotation periods. In \S3, I show that\clearpage \citet{Eggs} equilibrium tides combined with magnetic braking lead to different timescales for orbital decay, spin-orbit alignment, and retrograde flips, producing the observed trends. In \S4, I demonstrate via Monte Carlo that the tidal evolution defined in \S3 matches the observed trends and distributions. In \S5, I summarize the main features of the framework presented and discuss implications for discerning hot Jupiter migration mechanisms.

\section{Observed spin-orbit alignment trends}

I wish to explain: a) the host star effective temperature cut-off $\tcut$ for misaligned planets (W10), and b) the mass cut-off $\mret$ for retrograde planets \citep{Hebrard}. Fig. \ref{fig:obs} (left panels) displays $\lambda$ for hot Jupiters\footnote{\label{note:hj} Defined as $M_p>0.5M_{\rm Jup}$ and $P<7$ days. These criteria exclude three exceptions to the $\teff$ trend: low mass HAT-P-11-b \citep{2010W} and long period WASP-8-b \citep{2010Q} and HD-80606-b \citep{2009M}. WASP-80-b, a $0.5~M_{\rm Jup}$ planet orbiting a 4150~K star, would also be an exception, but its spin-orbit alignment remains under investigation \citep{2013T}. I exclude measurements that A12 characterizes as poorly-constrained: CoRoT-1, CoRoT-11, CoRoT-19, HAT-P-16, WASP-2, WASP-23, XO-2.  I only include stars with $\teff<6800$~K, which excludes KOI-13b.} and the host stars' projected stellar rotation frequency $\oms\sin~i_\star$ (computed from $v\sin~i_\star$, measured from spectral line broadening, and $R_\star$) versus $\teff$. (I plot $\oms\sin~i_\star$ instead of $v \sin i_\star$  for comparison with the simulations in \S3.) I note that, particularly below $\tcut$, stars hosting more massive planets have larger rotation frequencies (Fig. 1, bottom left). Due to the scarcity of massive hot Jupiters, the two trends relating to planetary mass are less robust than that with stellar temperature. Above $\tcut$, the chance occurrence that of the twelve significant obliquities, the three lowest correspond to the most massive planets is $\frac{12!3!}{15!}=0.2\%$; below $\tcut$, the chance occurrence that of the nineteen $\oms~\sin i_\star$, the two highest correspond to the most massive planets is $\frac{17!2!}{19!}=0.6\%$. These formal estimates of significance do not account for the variety of other patterns we might have observed\footnote{In the case of the mass stratification, I noticed the pattern in my simulations (\S3) before I examined the collection of observed $\oms$.}, which lowers their true significance.

As A12 showed (Fig. 20), because $\tcut$ for $\lambda$ is also a cut-off for $v\sin~i$ (or, plotted here, $\oms\sin~i$), it may correspond to the magnetic braking cut-off. This cut-off is equivalent to gyrochronological Kraft Break for stellar rotation period versus color \citep{Kraft}. Braking is thought to be strongest for stars with thick convective envelopes \citep{1962S}. Therefore A12 interpreted the correspondence of $\tcut$ for $\oms$ and $\psi$ as evidence that strong dissipation in the convective envelope plays a role in spin-orbit alignment. I will later argue that magnetic braking, rather than the tidal dissipation efficiency or the participation of the convective envelope versus entire star in tidal realignment, is the cause of $\tcut$ for $\psi$. 

I estimate the posterior distribution $\prob(\tcut)$ using a model with $\psi=0$ below $\tcut$ and isotropic above (following FW09 except assigning membership to the aligned versus isotropic population based on $\tcut$):
\begin{eqnarray}
\label{eqn:tcut}
\nonumber\\
\nonumber\\
\prob (\tcut) = \prod_i^{N \rm planets}(\prob(\lambda_i=0)\prob(\teff_i<\tcut)\nonumber \\
+\int_0^\pi~d\lambda_i~d\psi~\prob(\lambda_i|\psi) \nonumber \\ \prob(\psi)\prob\lambda_i\prob(\teff_i>\tcut)) \nonumber\\
\end{eqnarray}
\noindent for which $\prob(\lambda_i|\psi)$ is FW09 Eqn. 19, $\prob(\psi)=\frac{1}{2}\sin\psi$, and $\prob(\teff)$ and $\prob(\lambda_i)$ are normal distributions defined by their reported uncertainties. I plot $\prob(\tcut)$, peaking at 6090$^{+150}_{-110}~$K, and rotation rates measured by \citet{Mc} for $\sim10,000$ \kep targets (Fig. 2). The turnover in rotation rate is the Kraft Break; stars above this temperature remain rapidly rotating due to weaker magnetic braking. The Kraft Break matches the peak of $\prob(\tcut)$.

\section{Origin of the trends}

I hypothesize on the origin of the observed trends from the equations governing the planet's specific orbital angular momentum vector $\vec{h}$ [length$^2$/time] and the host star's spin angular frequency vector $\vec{\oms}$[time$^{-1}$] (\citealt{Eggs}, Eqn. 2--3). I assume the planet's orbit is circular, neglect terms that only cause orbital precession, and add a braking term (\citealt{1981V}, Eqn. 4; employed by \citealt{Barker} and W10). Equations \ref{eqn:h}--\ref{eqn:omegas} here correspond to \citet{Barker}, Eqn. A7 and A12 with the eccentricity vector $\vec{e}=0$.

\begin{equation}
\label{eqn:h}
\frac{\dot{\vec{h}}}{h}=-\frac{1}{\tau}\frac{\vec{h}}{h}+\frac{1}{\tau}\frac{\oms}{2n}(\frac{\vec{\oms}\cdot\vec{h}}{\oms~h}\cdot\frac{\vec{h}}{h}+\frac{\vec{\oms}}{\oms})\nonumber\\
\end{equation}
\begin{equation}
\label{eqn:omegas}
\dot{\vec{\oms}}=-\frac{M_p}{\ieff}\dot{\vec{h}}-\alphab\oms^2\vec{\oms}
\end{equation}
\noindent for which 
\begin{eqnarray}
\label{eqn:tau}
\tau=\frac{Q}{6k_L}\frac{M_\star}{R_\star^5(M_\star+M_p)^8G^7}\frac{M_\star}{M_p}h^{13} \nonumber \\=\tau_0\left(\frac{h}{h_0}\right)^{13}\frac{0.5M_{\rm~Jup}}{M_p}
\end{eqnarray} is an orbital decay timescale, $k_L$ is the Love number, $Q$ is the tidal quality factor, $M_p$ is the planet mass, $\ieff$ is the effective stellar moment of inertia participating in the tidal realignment, $\alphab$ is a braking constant, and $h_0 = \sqrt{a_0 G (M_\star+M_p)}$ is the initial specific angular momentum. In the simplified model here, $\tau_0$ is a constant. The timescale $\tau$ is related to \citet{Eggs} Eqn. 7 by replacing the viscous timescale with $Q$ (\citealt,Eqn. A10), altering the semi-major axis scaling from $a^8$ to $a^{13/2}$.
\begin{figure}
\includegraphics{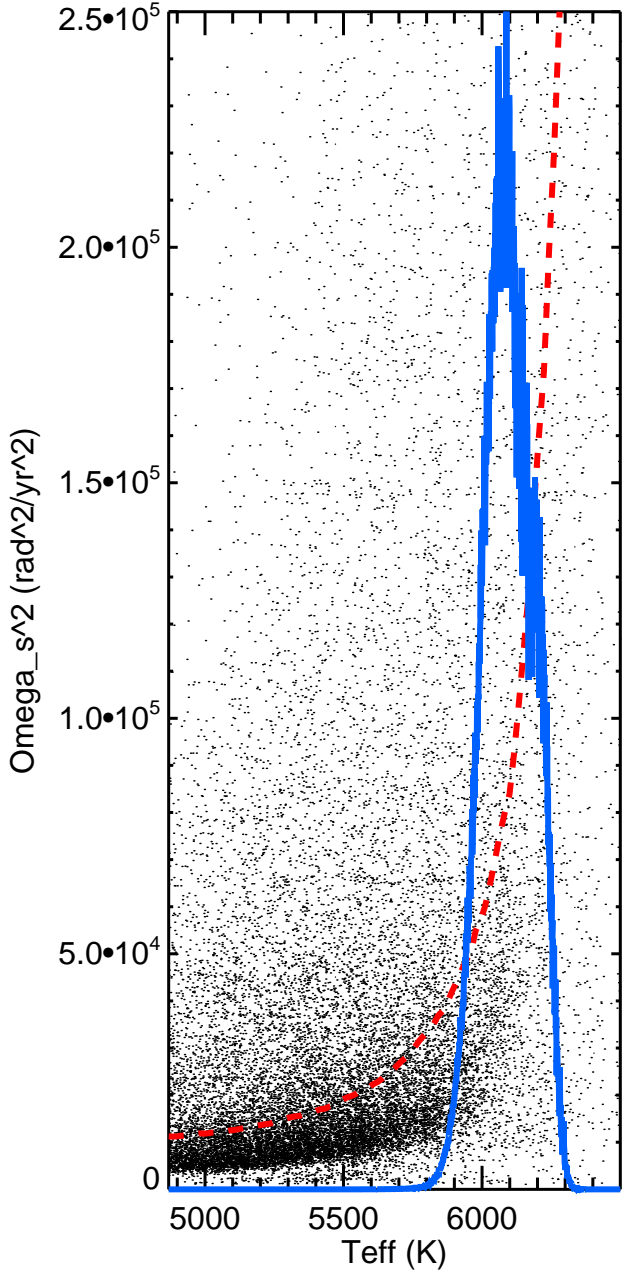}
\caption{Black dots: Squared stellar rotation rates $\oms^2$ ($0.8<M_\star<1.3M_\odot$) measured by \citet{Mc}. Red dashed line: Running median $\oms^2$, binning with 0.06 dex $\log_{10}\teff$. Blue solid line: $\prob(\tcut)$ from observed $\lambda$ (Eqn. \ref{eqn:tcut}). \label{fig:rot}}
\end{figure}

The spin-orbit alignment angle, $\cos \psi=\frac{(\vec{h}\times\vec{\oms})}{h\oms}$, evolves according to:
\begin{equation}
\label{eqn:psi}
\dot{\psi} =-\frac{1}{\talign}\sin\psi\left[1-\frac{\oms}{2n}\left(\cos\psi-\frac{\talign}{\tau}\right)\right]
\end{equation}
\noindent with characteristic alignment timescale
\begin{eqnarray}
\talign=\frac{\oms\ieff}{M_p h}\tau,
\end{eqnarray}
\noindent A sharp cut-off in alignment can be produced when $\psi$ accelerates. Neglecting the $\frac{\oms}{2n}$ term, when $h$ changes slowly compared to $\psi$ ($\tau>> \talign$), $\psi$ accelerates (very approximately) like:
\begin{equation}
\label{eqn:accel}
\frac{\ddot{\psi}}{\dot{\psi}}\sim\frac{1}{\tbrake}-\frac{\cos\psi}{\talign}
\end{equation}
\noindent for which 
\begin{equation}
\tbrake=\frac{1}{\alpha\oms^2},
\end{equation}
\noindent the first term dominates for cool stars, and --- for hot stars --- the second terms leads to quick flips for retrograde planets ($\cos\psi<0$) and gradual deceleration for prograde planets ($\cos\psi>0$). Thus $\tbrake$ and $\talign$, when compared to the orbital decay timescale $\tdec \sim \tau$ and the star's age ($\tage$; more precisely, the time since the hot Jupiter arrived at its close-in location), define tidal evolution regimes leading to the observed trends:
\begin{enumerate}
\item Misaligned regime ($\tdec,\tbrake,\talign>\tage$): $\vec{h}$ and $\vec{\oms}$ change very little, the {\bf regime for hot Jupiters with $M_p<\mret$ orbiting hot stars} (Fig. 1, black and blue triangles, $\teff>\tcut$)
\item Flipped regime ($\tdec,\tbrake>\tage>\talign$):  For planets massive enough to cause a short $\talign$, $\psi$ drops rapidly when the planet's orbit is retrograde ($\cos\psi<0,\frac{\ddot{\psi}}{\dot{\psi}}>0$), but falls off slowly when prograde ($\cos\psi>0,\frac{\ddot{\psi}}{\dot{\psi}}<0)$, allowing sufficiently massive planets to flip from a retrograde to prograde orbit, the {\bf regime for hot Jupiters with $M_p>\mret$ orbiting hot stars} (Fig. 1, red dots, $\teff>\tcut$)
\item Realigned regime  ($\tdec>\tage>{\talign}_0> \tbrake$): braking dominates the $\psi$ acceleration (Equation \ref{eqn:accel}), triggering a fast $\talign$ (due to a small $\oms$), the {\bf regime for hot Jupiters orbiting cool stars} (Fig. 1, $\teff<\tcut$)
\item Spin-down regime ($\tdec,\talign>\tage>\tbrake$): the star slows down but its spin orientation relative to the planet's orbit remains roughly constant $(\dot{\psi}\sim0)$, the {\bf expected regime for low mass planets orbiting cool stars}. HAT-P-11-b (footnote \ref{note:hj}, \citealt{2010W}) falls in this category.
\item Fast decay regime $(\tage>\tdec)$: the planet is consumed or tidally disrupted
\end{enumerate}
\noindent The $\talign/\tdec$ ratio is equivalent to the ratio of stellar spin to planetary orbital angular momentum, e.g., employed by \citet{Rogers13}.

Therefore I argue that the distinction between the hot versus cool stars is not the convective envelope versus entire star participating in the realignment but the host star's rotation rate due to magnetic braking. Although a smaller $\ieff$ for cool stars than for hot stars could in principle cause $\tcut$, hot stars also need a small $\ieff$ for regime 2. Therefore the distinction must be in $\oms$ (\S2), which differs systematically (hot stars spin quickly and cool stars slowly) due to a different magnetic braking strength. Furthermore, hot stars need weak magnetic braking for regime 2, whereas cool stars need strong magnetic braking for the planet to tidally realign the star's outer layer of  without synchronization (Fig. 1). Therefore I consider strong versus weak magnetic braking to be the cause of the observed trends, in concert with a small $\ieff$ for all stars, corresponding to an outer layer that participates in the tidal realignment while weakly coupled to the interior. Although in principle a small $\ieff$ is unnecessary if $\oms$ is initially extremely small, $\oms$ needs to match the observed $\oms\sin~i_s$ measurements (Fig. 1, bottom left panel). 

I plot several illustrative cases, using the parameters below (Fig. 3). The black dashed lines show a low-mass planet realigning a cool star. The red dashed lines represent a high-mass planet realigning a cool star.  The more massive planet spins up the star, resulting in a shorter stellar rotation period. The solid black lines depict a massive planet that orbits a hot star flipping from retrograde to prograde (regime 2), whereas the solid red lines represent a less massive planet inducing little realignment over the host star's lifetime (regime 1). In all cases, the planets experience little orbital decay (top panel).

\begin{figure}[h]
\includegraphics{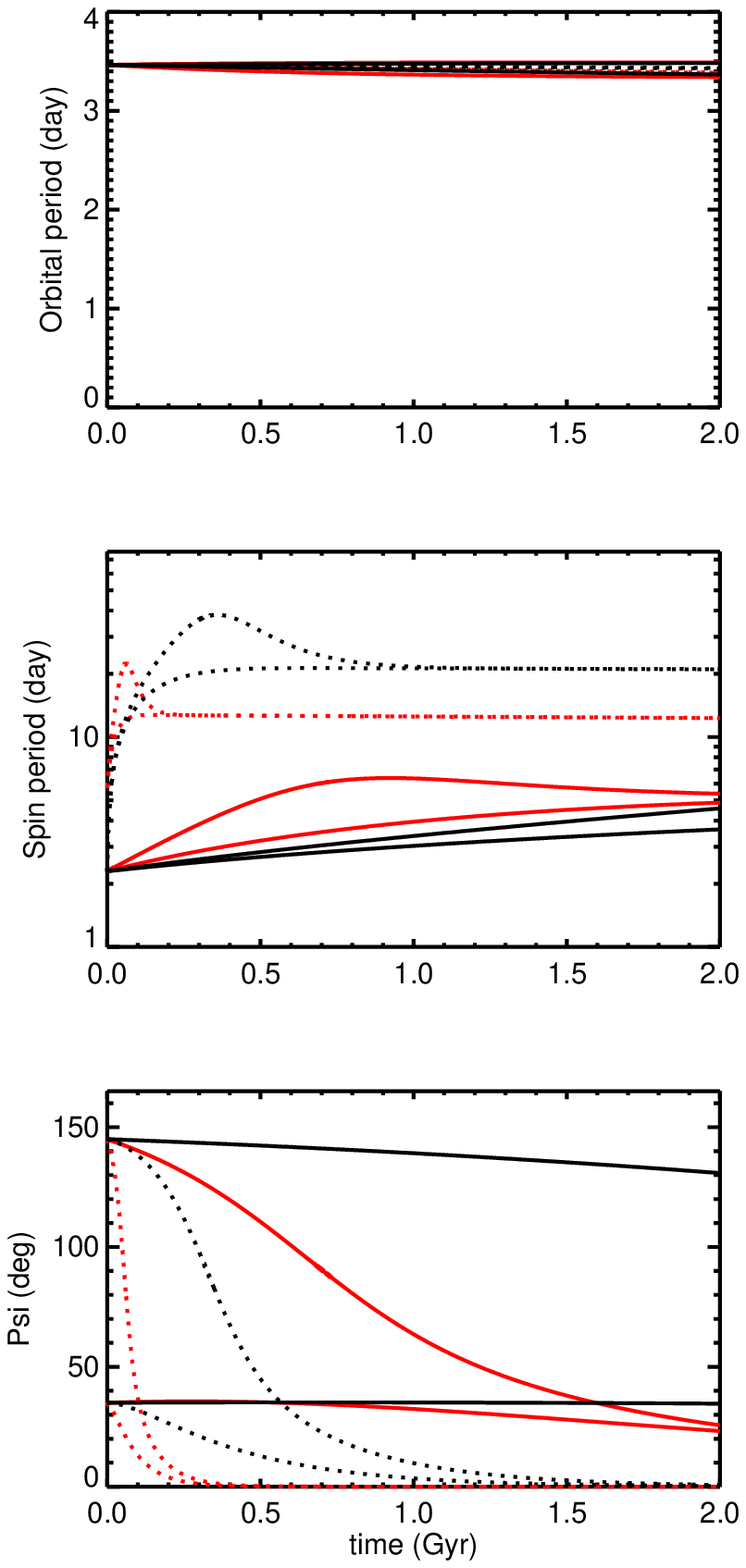}
\caption{Planetary $P$ (top), stellar spin period (middle), and $\psi$ (bottom) for $M_p=1M_{\rm~Jup}$ (black) and $M_p=3M_{\rm~Jup}$ (red) and hot (solid) or cool (dotted) host stars for an initial misalignment $\psi=145^\circ$ (corresponding to higher lines with bumps in middle panel) and $35^\circ$. \label{fig:example}}
\end{figure}

Using the timescales above, I estimate physical constants that produce the observed regimes.  For cool stars to realign and slow to the observed $(\oms)_{\rm~final}\sim225$ rad/yr, strong braking and low $\ieff$ are required:

\begin{eqnarray}
\label{eqn:cool}
\alpha_{\rm~cool}\sim3\times10^{-13}{\rm~yr}\left(\frac{225\rm~rad/yr}{(\oms)_{\rm~final}}\right)^2\frac{0.07\rm~Gyr}{t_{\rm~align}}\nonumber\\\nonumber\\\
\ieff\sim0.0004M_\odot~R_\odot^2\nonumber\\ 
\frac{t_{\rm~align}/t_{\rm~decay}}{0.003}\frac{M_p}{0.5M_{\rm~Jup}}\frac{h_0}{1.33\rm~AU^2/yr}\frac{225\rm~rad/yr}{(\oms)_{\rm~final}}, \nonumber\\
\end{eqnarray}
\noindent  much smaller than the sun's $I\sim(0.06M_\odot~R_\odot^2)$.

Next I estimate $\ieff$ for hot stars. For planets with $M_p>\mret$ to flip to prograde, I require:
\begin{eqnarray}
\label{eqn:hot}
\ieff\sim0.0019M_\odot~R_\odot^2\nonumber\\ 
\frac{\talign/t_{\rm~decay}}{0.007}\frac{\mret}{2.5M_{\rm~Jup}}\frac{h_0}{1.33\rm~AU^2/yr}\frac{600\rm~rad/yr}{(\oms)_{\rm~final}} \nonumber\\
\end{eqnarray}
\noindent Although this corresponds to a slightly larger $\ieff$ than for cool stars, it is still much smaller than the moment of inertia of the entire star (e.g., a $1.2~M_\odot,~1.2~R_\odot$ $n=3$ polytrope has $I=0.14M_\odot~R_\odot^2$). In \S\ref{sec:conclude}, I discuss whether such a small $\ieff$ for either hot or cool stars is plausible.

\section{Reproducing the observed trends}

I show via Monte Carlo that the framework from \S3 can reproduce the observed trends. I use the constants derived above (Eqn. \ref{eqn:cool}--\ref{eqn:hot}) and adopt $\tau_0=1000$ Gyr ($\tau$ scales with planet mass according to Eqn. \ref{eqn:tau}; $\tau_0$ corresponds to $Q\sim5\times10^6$ for a sun-like star), $(\oms)_{0,\rm~hot}=1000$ rad/yr, $(\oms)_{0,\rm~cool}=400$ rad/yr, $\alpha_{\rm~hot}=3\times10^{-16}$ yr, and $h_0=1.33$ AU$^2$/yr (corresponding to $P\sim3$ days, where the hot Jupiter pile-up is observed, e.g., \citealt{Gaudi}). The values are tuned to match the observations (Fig. 1).

To produce a predicted population, for 200 planets I:
\begin{enumerate}
\item Select a uniform random  $4800<\teff<6800~$K, a log-uniform $0.5M_{\rm~Jup}<M_p<25M_{\rm~Jup}$, and $\psi$ from an isotropic distribution.
\item Select a uniform random evolution time $0<t<10$ Gyr for cool stars ($\tcut<6100$K) or $0<t<4$ Gyr for hot stars. Integrate Eqn. \ref{eqn:h}, \ref{eqn:omegas}, and \ref{eqn:tau} for $t$. 
\item Compute $\psi(t)$ and select a uniform random longitude of ascending node $0<\Omega<2\pi$, yielding the sky-projected $\lambda=\tan^{-1}\left(\tan\psi\sin\Omega\right)$ (FW09, Eqn. 11).
\item Compute $\sin~i_\star=\sqrt{1-(\sin\psi\cos\Omega)^2}$ and $v\sin~i_\star/R_\star=\oms\sin~i_\star$.
\end{enumerate}

Fig. 1 (right panels) displays simulated planets not subsumed by their stars (most have $P$ close to the initial value). Qualitatively I match the observed distributions and trends in $\lambda$ and $\oms$ (Fig. 1, left panels): the sharp transition at $\tcut$ between aligned and misaligned planets, the mass cut-off for retrograde planets, the larger rotation frequencies for hot stars, and the planet mass stratification of the projected rotation frequencies for cool stars. The mass stratification results from the highest mass planets spinning up their host stars, counteracting magnetic braking.

I repeat the simulation replacing $Q$ with a viscous time constant, such that $\tau\propto~h^{16}$: the results do not change noticeably, likely because the planet experiences little orbital decay. I can increase $\ieff$ for cool stars to $0.004M_\odot~R_\odot$ by decreasing $\tau_0$ to 100 Gyr for cool stars only, but more planets are tidally disrupted. I cannot more than double $\ieff$ for hot stars. The results are not strongly sensitive to the initial $P$, which mostly affects how much the massive planets spin up their host stars; a longer $P$ also, via the $\frac{\oms}{2n}$ term (Eqn. \ref{eqn:psi}), partially counteracts the realignment. I try to reproduce the observed trends and distributions in several alternative regimes (not shown): large $\ieff$ with strong braking or low initial $\oms$, large $\ieff$ with initially strong braking that drops after 100 Myr, and weak braking for cool stars but smaller initial $\oms$. I find that although I can reproduce the distribution and trends in $\lambda$ in these regimes, I cannot simultaneously reproduce the observed $\oms\sin~i_s$.

\section{Conclusion}
\label{sec:conclude}
      
I advocate that equilibrium tides with magnetic braking and a small effective stellar moment of inertia participating in the tidal realignment ($\ieff$) can account for the observed trends and distributions of hot Jupiters' spin-orbit alignments and host star projected rotation frequencies. This framework is based on W10 and A12 but with several modifications to match the observations. The observed temperature cut-off between aligned and misaligned planets is due to stellar magnetic braking, rather than the tidal dissipation efficiency or whether the star's convective envelope is tidally realigned independent of the interior. For cool stars, strong magnetic braking\footnote{Magnetic braking was included in W10's realignment example and highlighted by W10 as enabling realignment without synchronization.}, slows the rotation and lowers the spin angular momentum, allowing the planet to realign the star's outer layer without synchronization. Both hot and cool stars require a $\ieff$ 30-100 times lower than the star's total $I$. For cool stars, magnetic braking shrinks the star's spin vector without changing the direction, so a small $\ieff$ allows a smidgen of orbital decay to nudge the star to realignment. For hot stars, a small $\ieff$ allows sufficiently massive planets to flip from retrograde to prograde, producing the observed retrograde planetary mass cut-off. Finally, more massive planets partially overcome magnetic braking to spin up the star, leading to the observed planet-mass-stratification of host star rotation frequency for cool stars. 

There are several caveats. The observed trends with planet mass are less robust than that with stellar effective temperature due to the rarity of massive hot Jupiters. Therefore Rossiter-McLaughlin measurements of newly discovered hot Jupiters with $M_p>2.5M_{\rm Jup}$ would be particularly valuable. I assumed that the initial $\psi$ distribution is independent of $M_p$; if the retrograde mass-cut were caused by the mechanism leading to misalignment, a small $\ieff$ for hot stars would not be necessary. Furthermore, whether such a small $\ieff$ --- an outer layer participating in the tidal realignment while weakly coupled to the interior --- is plausible remains an open question. As proposed by W10, this weakly coupled outer layer could be the convective zone. Following W10, I compute the convective zone $I$ of a 5 Gyr sun-like star and a 2 Gyr 1.2 solar-mass star using the EZ Web stellar evolution models \citep{2004P}: $0.01~M_\odot~R_\odot^2$ and $0.002M_\odot~R_\odot^2$ respectively. The latter is consistent with the value estimated here ($0.0019~M_\odot~R_\odot^2$). The former is inconsistent with the estimated value $0.0004M_\odot~R_\odot^2$ using nominal parameters and marginally consistent with $0.004~M_\odot~R_\odot^2$ if I allow cool stars to have a much shorter orbital decay timescale. The weakly coupled outer layer does not necessarily correspond to the convective zone, just an effective distance for the transferred angular momentum to penetrate. A weakly coupled outer layer would account $\tau$ Bootis~b synchronizing its hot star. For cool stars, a weakly coupled outer layer seems at odds with the sun's radially uniform rotation profile. One possibility is that the weakly coupled outer layer corresponds to our sun's near-surface shear outer layer at about 0.95$~R_\odot$ (e.g., \citealt{1996T}), which is decoupled from the rest of the convective zone at low latitudes. The depth from the surface probed by  $\lambda$ and $v\sin~i_\star$ measurements is $<100$ km (e.g., \citealt{Gray}). Another possibility is that the timescale for coupling between layers, often treated as a free parameter in tidal evolution models due to uncertainty over which proposed physical process is responsible for coupling (e.g., \citealt{Allain,Penev}), is much longer than the tidal forcing timescale.

I used a simplistic tidal evolution model to illuminate an origin for the observed stellar obliquity trends. Ultimately a detailed, realistic treatment is necessary, incorporating stellar evolution (accounting for age trends, i.e. \citealt{2011T}), coupling between the star's layers, changes in $\alphab$, dynamic tides, stellar properties affecting tides (e.g., $R_\star$, and tides raised on the planet. The values of physical constants I tuned to match observations--- such as the braking coefficient and initial host star spin rate --- are not precise constraints due to the model's simplistic treatment, but better models may allow meaningful constraints. For example, a constraint on the initial host star spin rate could pinpoint the star's age when the realignment begins, potentially distinguishing between disk migration (which occurs before the gas disk evaporates) versus high eccentricity migration (which delivers the hot Jupiter over a larger range of timescales).

A key take-away for using the $\psi$ distribution to constrain hot Jupiter migration mechanisms is that cool star obliquities are significantly sculpted by tides, as are hot star obliquities for the more massive hot Jupiters. For example, if one were to take the $\psi$ distribution of hot stars as pristine, one would overestimate the fraction of prograde planets due to the flipping of massive retrograde planets. Better tidal evolution models will allow robustly forward-modeling the migration and tidal sculpting process to match the observed distribution of sky-projected $\lambda$.

\acknowledgments
My gratitude to the referee for an especially helpful report. I thank Joshua Winn, Gwena{\"e}l Bou{\'e}, Eliot Quataert, Daniel Fabrycky, Ryan O'Leary, Eugene Chiang, Simon Albrecht, Amaury Triaud, Marta Bryan, Howard Isaacson, Ruth Murray-Clay, and Ian Czekala for helpful feedback, comments, and discussions, Ruth Angus for an inspiring presentation on gyrochronology, the Miller Institute for Basic Research in Science, University of California Berkeley for funding, and the Exoplanet Orbit Database ({\tt exoplanets.org}) .

\end{document}